\documentclass[letterpaper,journal]{IEEEtran}
\usepackage{amsmath,amsfonts}
\usepackage{bm}
\usepackage{algorithmic}
\usepackage{algorithm}
\usepackage{array}
\usepackage[caption=false,font=normalsize,labelfont=sf,textfont=sf]{subfig}
\usepackage{textcomp}
\usepackage{stfloats}
\usepackage{url}
\usepackage{verbatim}
\usepackage{graphicx}
\usepackage{cite}
\usepackage{color}
\usepackage{tikz,xcolor}
\usepackage[implicit=false]{hyperref}
\hypersetup{hidelinks,
	colorlinks=true,
	allcolors=black,
	pdfstartview=Fit,
	breaklinks=true}

\begin{document}

\title{Symbol-wise Puncturing for HARQ Integration \\ with Probabilistic Amplitude Shaping}

\author{Li Shen,
        ~Yongpeng Wu,~\IEEEmembership{Senior~Member,~IEEE},
        ~Derrick Wing Kwan Ng,~\IEEEmembership{Fellow,~IEEE},
        \\Wenjun Zhang,~\IEEEmembership{Fellow,~IEEE},
        ~and~Xiang-Gen Xia,~\IEEEmembership{Fellow,~IEEE}
\thanks{L. Shen, Y. Wu, and W. Zhang are with the Department of Electronic Engineering, Shanghai Jiao Tong University, Shanghai 200240, China (e-mail: \href{mailto:shen-l@sjtu.edu.cn}{shen-l@sjtu.edu.cn}; \href{mailto:yongpeng.wu@sjtu.edu.cn}{yongpeng.wu@sjtu.edu.cn}; \href{mailto:zhangwenjun@sjtu.edu.cn}{\mbox{zhangwenjun}@sjtu.edu.cn}) \emph{(Corresponding author: Yongpeng Wu.)}.}
\thanks{D. W. K. Ng is with the School of Electrical Engineering and Telecommunications, University of New South Wales, Sydney, NSW 2052, Australia (e-mail: \href{mailto:w.k.ng@unsw.edu.au}{w.k.ng@unsw.edu.au}).}
\thanks{X.-G. Xia is with the School of Information, and Electronics, Beijing Institute of Technology, Beijing 100081, China, and also with the Department of Electrical, and Computer Engineering, University of Delaware, Newark, DE 19716 USA (e-mail: \href{mailto:xianggen@udel.edu}{xianggen@udel.edu}).}
}



\maketitle

\begin{abstract}
In this letter, we propose a symbol-wise puncturing scheme to support hybrid automatic repeat request (HARQ) integrated probabilistic amplitude shaping (PAS). To prevent the probability distribution distortion caused by the traditional sequential puncturing and realize the promised gain of PAS, we perform symbol-wise puncturing on the label sequence of the shaped modulation symbols. Our simulation results indicate that the proposed puncturing scheme achieves a stable shaping gain across the signal-to-noise ratio of at least 0.6 dB compared with the uniform case under the same throughput, while the gain of sequential puncturing drops rapidly with retransmissions. Moreover, in each transmission, the proposed scheme is able to reduce the distribution distortion that achieves over 1.2 dB gain at a block error rate (BLER) of $\bm{10^{-3}}$. In contrast, for sequential puncturing, the distribution is severely distorted and the BLER performance is even worse than that of the uniform case in retransmissions.
\end{abstract}


\section{Introduction}
\IEEEPARstart{T}{he} Shannon limit defines the maximum theoretical transmission rate for reliable communication over the additive white Gaussian noise (AWGN) channel \cite{shannon1948mathematical}. Yet, the limit is only achievable with Gaussian inputs and it remains a gap up to 1.53 dB with traditional uniformly distributed discrete input constellations \cite{forney1984efficient}, such as bipolar amplitude-shift keying (ASK) constellations and quadrature amplitude modulation (QAM) constellations. Therefore, numerous researchers have devoted to work on probabilistic shaping (PS), which converts the discrete input distribution into a Gaussian-like shape to bridge the shaping gap. For instance, Gallager first proposed PS implemented by the approach of many-to-one mapping in \cite{gallager1968information}. Unfortunately, performing many-to-one demapping at the receiver with low complexity is challenging when a binary forward error correction (FEC) code is applied. Futhermore, other PS schemes, such as trellis shaping\cite{forney1992trellis}, shell mapping \cite{khandani1993shaping}, and concatenated shaping \cite{le2005bit}, also have common shortcomings of computational complexity and rate inflexibility. As a remedy, the authors in \cite{bocherer2015bandwidth} proposed the probabilistic amplitude shaping (PAS) scheme with low implementation complexity and transmission rate flexibility. In this scheme, a distribution matcher (DM) is used for PS that can be seamlessly integrated with existing FEC. As at the receiver, it performs bit-metric decoding \cite{bocherer2014achievable} without requiring any iterative demapping. Moreover, by adjusting the probability distribution, the PAS scheme flexibly adapts to different transmission rates. Numerical results in \cite{bocherer2015bandwidth} have shown that the PAS scheme achieves a gap performance less than $1.1$ dB to the AWGN capacity by using the ASK constellations and DVB-S2 low-density parity-check (LDPC) codes.


In addition to pursuing a high transmission rate, it is essential to provide reliable transmission in wireless communications systems. Therefore, the retransmission-based automatic repeat request (ARQ) schemes and in particular hybrid-ARQ (HARQ) schemes, which combines ARQ and FEC, has been widely adopted for error correction in practical systems. Generally, there are mainly three HARQ types according to the difference of data retransmitted \cite{ahmadi2013lte}: type-\uppercase\expandafter{\romannumeral1}, type-\uppercase\expandafter{\romannumeral2} and type-\uppercase\expandafter{\romannumeral3}. Specifically, type-\uppercase\expandafter{\romannumeral1} HARQ is also denoted as chase combining HARQ (CC-HARQ), in which the same data packet is transmitted for all retransmissions. As such, its performance mainly depends on the error correction ability of FEC. Unlike CC-HARQ, type-\uppercase\expandafter{\romannumeral2} HARQ, also denoted as incremental redundancy HARQ (IR-HARQ), only transmits the redundant information whenever they are needed to bring higher throughput. On the other hand, Type-\uppercase\expandafter{\romannumeral3} HARQ is similar to IR-HARQ but each data packet is required to be self-decodable, which means that the information bits can be extracted independently of other transmitted data packets.

In recent years, the PAS scheme has been widely investigated in wireless and fiber-optic communications, e.g. \cite{gultekin2020enumerative,bocherer2019probabilistic,matsumine2022rate}. As a key component of PAS, the design of DM has also attracted much attention\cite{fehenberger2020huffman,fu2021parallel}. Meanwhile, the researches on HARQ mainly focus on more efficient retransmission process, such as machine learning-based\cite{ceran2021reinforcement} and non-orthogonal\cite{nadeem2021nonorthogonal} approaches. However, none of the existing works have considered the integration between PAS and HARQ.

In this letter, we aim to integrate PAS with HARQ. When applying IR-HARQ, simply performing the sequential puncturing may cause potential distribution distortion in retransmissions. As such, we propose a symbol-wise puncturing scheme to preserve the shaping distribution and realize the promised gain of PAS. First, we consider the standard PAS structure\cite{bocherer2015bandwidth} and adopt the constant composition distribution matcher (CCDM) proposed in \cite{schulte2015constant} as a DM to transform the input bits into amplitude symbols. Then, they are represented by their binary labels with a desired probability distribution, while the binary labels are termed as \emph{amplitude bits}. After FEC encoding, the information bits excluding the amplitude bits and all the parity bits are called \emph{sign bits}. We apply IR-HARQ in our work, i.e., transmitting data packet that is self-decodable for the first transmission and the redundant information for retransmissions. In this case, the traditional sequential puncturing will cause retransmissions with only the sign bits, which brings severe distribution distortion. To address this issue, we pair the amplitude bits and sign bits into a label sequence of shaped modulation symbols. Then, we perform symbol-wise puncturing on the label sequence. In our scheme, a certain proportion of the amplitude bits and sign bits are sent in each transmission to ensure the probability distribution. At the receiver, the a-priori information of the amplitude bits that have not be transmitted is calculated to enhance FEC decoding. Our simulation results indicate that our proposed symbol-wise puncturing can obtain stable shaping gain across the signal-to-noise ratio (SNR) on the throughput over both AWGN and multiple-input multiple-output (MIMO) Rayleigh fading channels and realize the desired probability distribution with only little distortion in each transmission, while the gain of the sequential puncturing drops rapidly with the increase of retransmission times and the distribution of the sequential puncturing in retransmissions tends to be uniform. Note that for type-\uppercase\expandafter{\romannumeral1} and type-\uppercase\expandafter{\romannumeral3} HARQ, since the data packets in each transmission are self-decodable, there are enough amplitude bits to preserve the probability distribution. In other words, the conventional PAS would still work when it is integrated with these two types of HARQ. Therefore, we only consider IR-HARQ in our work.

\section{System Model}
\begin{figure}[!t]
        \centering
        \includegraphics[scale=0.72]{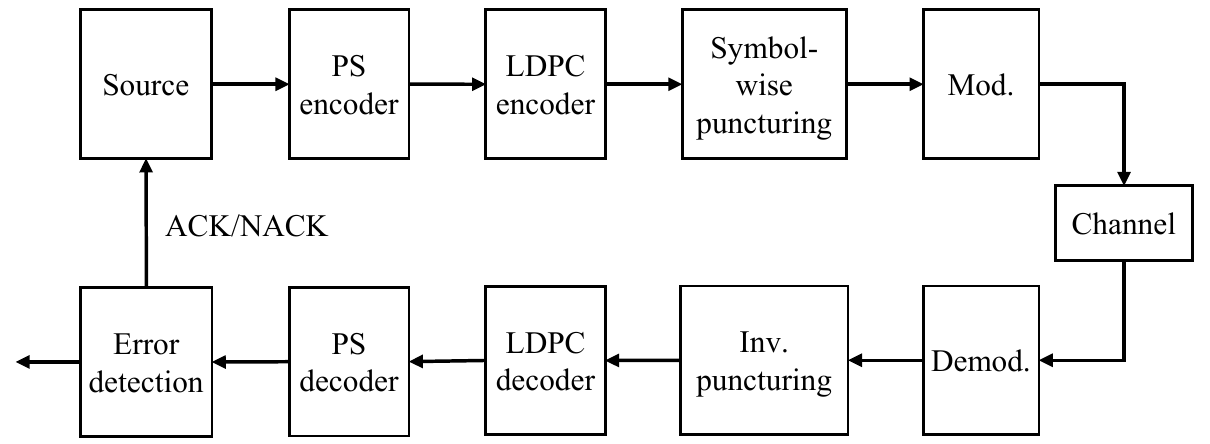}
        \caption{System model of symbol-wise puncturing with PAS and HARQ.}
        \label{fig:1}
\end{figure}

The block diagram of the proposed symbol-wise puncturing system with PAS and HARQ is shown in Fig.~\ref{fig:1}. For the first transmission, the source generates a random binary data block. Then, the data block is mapped to the shaped bits by a PS encoder while an LDPC encoder is performed for FEC encoding subsequently. After symbol-wise puncturing the encoded codeword, the punctured bits are modulated into QAM symbols and transmitted over the channel. On the other hand, at the receiver, the soft demodulator adopts the received noisy symbols to calculate the log-likelihood ratios (LLRs). Then, the received encoded codeword is recovered by inverse puncturing and feed the intermediate results to the LDPC decoder. Subsequently, the PS decoder transforms the decoded bits back to the data block. If there is no error, the error detection returns an acknowledgment (ACK) and the transmitter sends the next data block. However, if the data block cannot be correctly decoded, it returns a negative acknowledgment (NACK) to request a retransmission. Futhermore, if the data cannot be successfully recovered after reaching the maximum number of transmissions, it is dropped and an ACK is fed back.

\begin{figure}[!t]
        \centering
        \includegraphics[scale=0.85]{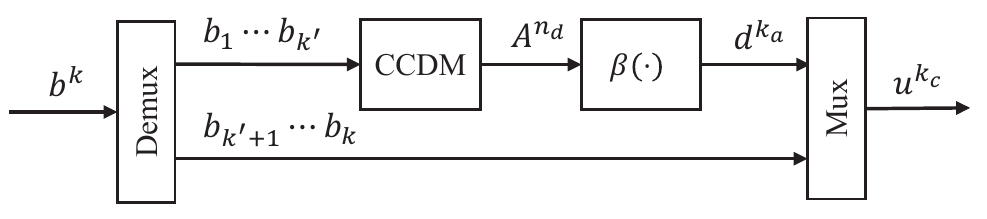}
        \caption{The PS encoder.}
        \label{fig:2}
\end{figure}

To combine PAS with LDPC and QAM modulation, the component of the PS encoder is detailed in Fig.~\ref{fig:2} \cite{bocherer2019probabilistic}. A binary sequence $b^k=b_1 b_2\cdots b_k$ is split into two sequences. The first $k^\prime$ bits are transformed to $n_d$ amplitude symbols $A^{n_d}=A_1 A_2\cdots A_{n_d}$ by a DM with rate $R_\text{dm} = k^\prime/n_d$ ~bit/symbol. For the $M^2$-QAM modulation scheme, the output amplitude symbols, $A_i\in\mathcal{A} = \{1,3,\cdots,M-1\},~i=1,\dots,n_d$, obey the desired probability distribution $P_A$. Then, $A^{n_d}$ is mapped to bit sequence $d^{k_a}$. For a QAM symbol carrying $m=\log_2 M^2$ bits, we require that the two sign bits label the quadrant of one symbol, while the remaining $m-2$ amplitude bits label the symbol's position in the quadrant. Therefore, after distribution matching, every two amplitude symbols $A_i A_{i+1}$, which correspond to the in-phase and quadrature components of the QAM constellation point in the first quadrant, respectively, are mapped to $m-2$ amplitude bits $\beta(A_i A_{i+1})$. As such, $d^{k_a}$ satisfies $d^{k_a}=\beta(A_1A_2)\cdots\beta(A_{n_d-1}A_{n_d})$ and $k_a = n_d(m-2)/2$. Finally, the PS encoder outputs the $k_c$ shaped bits $u^{k_c}$ combined by $d^{k_a}$ and the rest of $b^k$, i.e., $u^{k_c}=d_1 \cdots d_{k_a} b_{k^\prime+1} \cdots b_k$. Note that as CCDM is invertible~\cite{schulte2015constant}, the PS decoder can be realized by the inverse process of the encoder.

\begin{figure}[!t]
        \centering
        \includegraphics[scale=0.8]{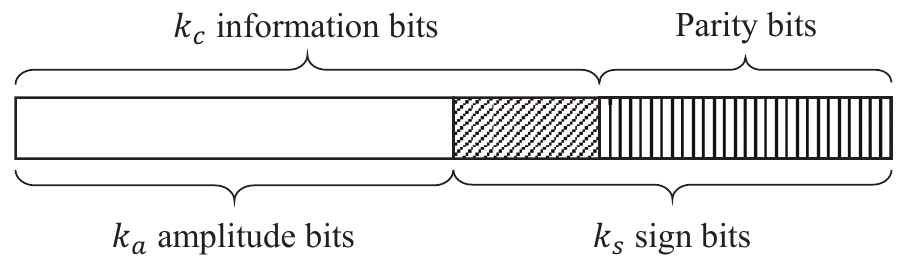}
        \caption{The structure of the shaped LDPC codeword.}
        \label{fig:3}
\end{figure}

After passing the LDPC encoder, the shaped bits $u^{k_c}$ are encoded to a codeword $c^{n_c}=c_1c_2\cdots c_{n_c}$ with code rate $R = k_c/n_c$. Fig.~\ref{fig:3} shows the codeword structure. $k_s$ sign bits consist of bit sequence $b_{k^\prime+1} \cdots b_k$ and parity bits.  And $k_s$ must be constrained by $k_a/k_s=(m-2)/2$ to be compatible with QAM modulation.

\section{Symbol-wise Puncturing}
\begin{figure}[!t]
        \centering
        \includegraphics[scale=0.8]{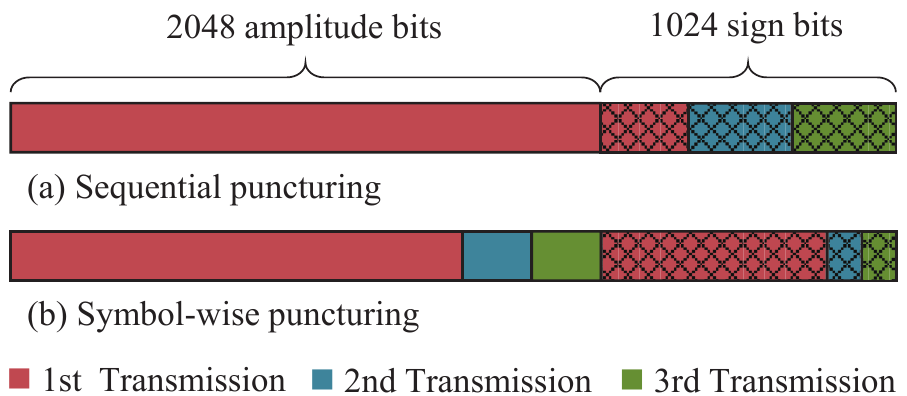}
        \caption{Examples of (a) sequential puncturing and (b) symbol-wise puncturing with $t_\text{max} = 3$, $n=512, n_1=392,n_2=n_3=60$, and $m=6$.}
        \label{fig:4}
\end{figure}

As discussed in Section~\uppercase\expandafter{\romannumeral2}, if we apply the traditional sequential puncturing to the shaped codeword shown in Fig.~\ref{fig:3}, the distribution of modulated symbols is distorted in retransmissions as no amplitude bits are sent. Therefore, we propose the symbol-wise puncturing. The basic idea of the symbol-wise puncturing is to pair the two sign bits and the $m-2$ amplitude bits into the label of a shaped symbol before puncturing and then perform puncturing in labels to realize the desired distribution. Assuming that there are totally $n=n_c/m$ symbols to be transmitted and the maximum number of transmissions is $t_\text{max}$, at the $t$-th transmission, we aim to transmit $n_t$ symbols subject to the constraint $\sum_{t=1}^{t_\text{max}} n_t=n$. First, a length-$m$ bit string, $\bm{S}_i$, is defined as
\begin{equation}
        \setlength{\abovedisplayskip}{4pt}
        \setlength{\belowdisplayskip}{4pt}
        \label{eq:1}
        \bm{S}_i = c_{k_a+2i-1}c_{k_a+2i}\beta(A_{2i-1}A_{2i}),
\end{equation}
where $\bm{S}_i$ actually labels a shaped symbol $x\in \mathcal{X}$ and $\mathcal{X}$ denotes $M^2$-QAM constellation. Note that the position of $\beta(A_{2i-1}A_{2i})$ in $\bm{S}_i$ varies with different modulation mappers. For convenience, we assume that the first two bits are sign bits and the last $m-2$ bits are the amplitude bits here. Then the symbol-wise punctured bit sequence for $t$-th transmission is given by
\begin{equation}
        \setlength{\abovedisplayskip}{4pt}
        \setlength{\belowdisplayskip}{4pt}
        \label{eq:2}
        \bm{S}_{p_t}^{q_t} = \bm{S}_{p_t} \bm{S}_{p_t+1} \cdots \bm{S}_{q_t},
\end{equation}
where $p_t = \sum_{j=1}^{t-1} n_j +1$ and $q_t = \sum_{j=1}^{t} n_j$. Fig.~\ref{fig:4} shows an example for $t_\text{max} = 3$. Unlike sequential puncturing that transmits all the amplitude bits at the first transmission, our symbol-wise puncturing only transmits partial amplitude bits and keep the ratio of amplitude bits to sign bits as $(m-2)/m$ per transmission. Thus, $\bm{S}_{p_t}^{q_t}$ can be directly modulated into symbols with desired distribution.

\subsection{Minimum Length of the First Transmission}
When applying the symbol-wise puncturing, there is a constraint on the symbol length of the first transmission. Since we require the amplitude bits to satisfy $k_a = n_c(m-2)/m$, the length of information bits must be greater than $k_a$ and the LDPC code rate satisfies
\begin{equation}
        \setlength{\abovedisplayskip}{4pt}
        \setlength{\belowdisplayskip}{4pt}
        R = \frac{k_c}{n_c} \geq \frac{m-2}{m}.
\end{equation}
If the receiver successfully recovers the data block after the $t$-th transmission, the transmission code rate is increased to
\begin{equation}
        \setlength{\abovedisplayskip}{4pt}
        \setlength{\belowdisplayskip}{4pt}
        \label{eq:3}
        R_t = \frac{k_c}{m\sum_{j=1}^t n_j}, 1\leq t \leq t_\text{max}.
\end{equation}
Then, the code rate $R_t$ is bounded by
\begin{equation}
        \setlength{\abovedisplayskip}{4pt}
        \setlength{\belowdisplayskip}{4pt}
        \frac{k_c}{n_c} \leq R_t < 1, 1\leq t \leq t_\text{max}.
\end{equation}
For $t=1$,
\begin{equation}
        \setlength{\abovedisplayskip}{4pt}
        \setlength{\belowdisplayskip}{4pt}
        \label{eq:4}
        \frac{k_c}{m\cdot n_1} < 1.
\end{equation}
As for $t=t_\text{max}$,
\begin{equation}
        \setlength{\abovedisplayskip}{4pt}
        \setlength{\belowdisplayskip}{4pt}
        \label{eq:5}
        \frac{k_c}{m\cdot n} \geq \frac{k_c}{n_c} = \frac{(m-2)n+k-k^\prime}{m\cdot n}.
\end{equation}
Combining (\ref{eq:4}) and (\ref{eq:5}), we obtain that $n_1$ is limited by
\begin{equation}
        \setlength{\abovedisplayskip}{4pt}
        \setlength{\belowdisplayskip}{4pt}
        \label{eq:6}
        n_1 > n\cdot \left( \frac{m-2}{m} + \frac{k-k^\prime}{m\cdot n} \right),
\end{equation}
which bounds the minimum number of symbols for the first transmission from below. 

\subsection{Inverse Puncturing}
At the receiver, the soft demodulator calculates the LLRs and outputs the estimated bit sequence $\hat{\bm{S}}_{p_t}^{q_t}$ for $t$-th transmission. For the inverse puncturing, the previous $t$ transmissions are combined into $\hat{\bm{S}}_{p_1}^{q_t}$. Then, $\hat{\bm{S}}_{p_1}^{q_t}$ is transformed to the estimated codeword $\hat{c}^{n_c}$ for LDPC decoding by the inverse process of (\ref{eq:1}). Usually, the non-transmitted bits are filled with 0 in $\hat{c}^{n_c}$. Yet, for the amplitude bits that have not been transmitted, we replace them with their a-priori information rather than 0 for better decoding performance. Let the bit string $\bm{B}=B_1B_2\cdots B_m$ with probability distribution $P_{\bm{B}}$ on $\{0,1\}^m$ denotes the label of symbol $x\in\mathcal{X}$. The a-priori information for $j$-th bit is calculated by
\begin{equation}
        \setlength{\abovedisplayskip}{4pt}
        \setlength{\belowdisplayskip}{4pt}
        L_{A,j} = \log\frac{P_{B_j}(0)}{P_{B_j}(1)} = \log \frac{ \sum_{\bm{a}\in\{0,1\}^m:a_j=0}P_{\bm{B}}(\bm{a}) }{ \sum_{\bm{a}\in\{0,1\}^m:a_j=1}P_{\bm{B}}(\bm{a}) }.
\end{equation}
As $B_3\cdots B_m$ correspond to the amplitude bits, we can obtain the a-priori information for $\beta(A_i A_{i+1})$ by calculating $L_{A,3},\dots,L_{A,m}$. Note that for each $\beta(A_i A_{i+1})$, we have the same a-priori information. Thus, the inverse puncturing eventually outputs estimated codeword $\hat{c}^{n_c}$ combined with a-priori information of untransmitted amplitude bits.

\subsection{HARQ Process}
After introducing our symbol-wise puncturing and inverse puncturing, an HARQ process with symbol-wise puncturing and PAS works as follows.
\begin{enumerate}
\item {Determine the maximum transmission times $t_\text{max}$. Initialize $t=1$. Generate a new data block. Perform PS encoding and LDPC encoding.}
\item {At the $t$-th transmission, obtain symbol-wise punctured bit sequence $\bm{S}_{p_t}^{q_t}$ according to (\ref{eq:2}).}
\item {Modulate $\bm{S}_{p_t}^{q_t}$ to QAM symbols. Then, the symbols are transmitted over a channel and received by a soft demodulator.}
\item {Perform inverse puncturing, which is detailed in Section~\uppercase\expandafter{\romannumeral3}-B.}
\item {Perform LDPC decoding and PS decoding.}
\item {If the data block is successfully recovered or $t$ reaches $t_\text{max}$, return an ACK and end this transmission. If not, return a NACK to request a retransmission, let $t=t+1$ and go back to step 2.}
\end{enumerate}

\section{Simulation Results}
In this section, we evaluate the performance of the proposed symbols-wise puncturing. The LDPC codes are from the WLAN standard \cite{WIFI5} and are decoded by the belief-propagation~(BP) algorithm with a maximum iteration number of 12. The output probability distribution $P_A$ of CCDM can be obtained by \cite[Sec.~\uppercase\expandafter{\romannumeral3}]{bocherer2015bandwidth} and obeys the Maxwell-Boltzmann (MB) distribution. We choose CCDM rate as $R_\text{dm} = \mathbb{H}(A)$, where $\mathbb{H}(\cdot)$ denotes the entropy function. Note that when $R_\text{dm}$ is fixed, since the LDPC code rate provided is not continuous, we may need to append some filler bits (zero bits), which are assumed to be known at both the transmitter and the receiver, to reach the length of information bits after PS encoding.

\begin{figure}[!t]
        \centering
        \includegraphics[scale=0.6]{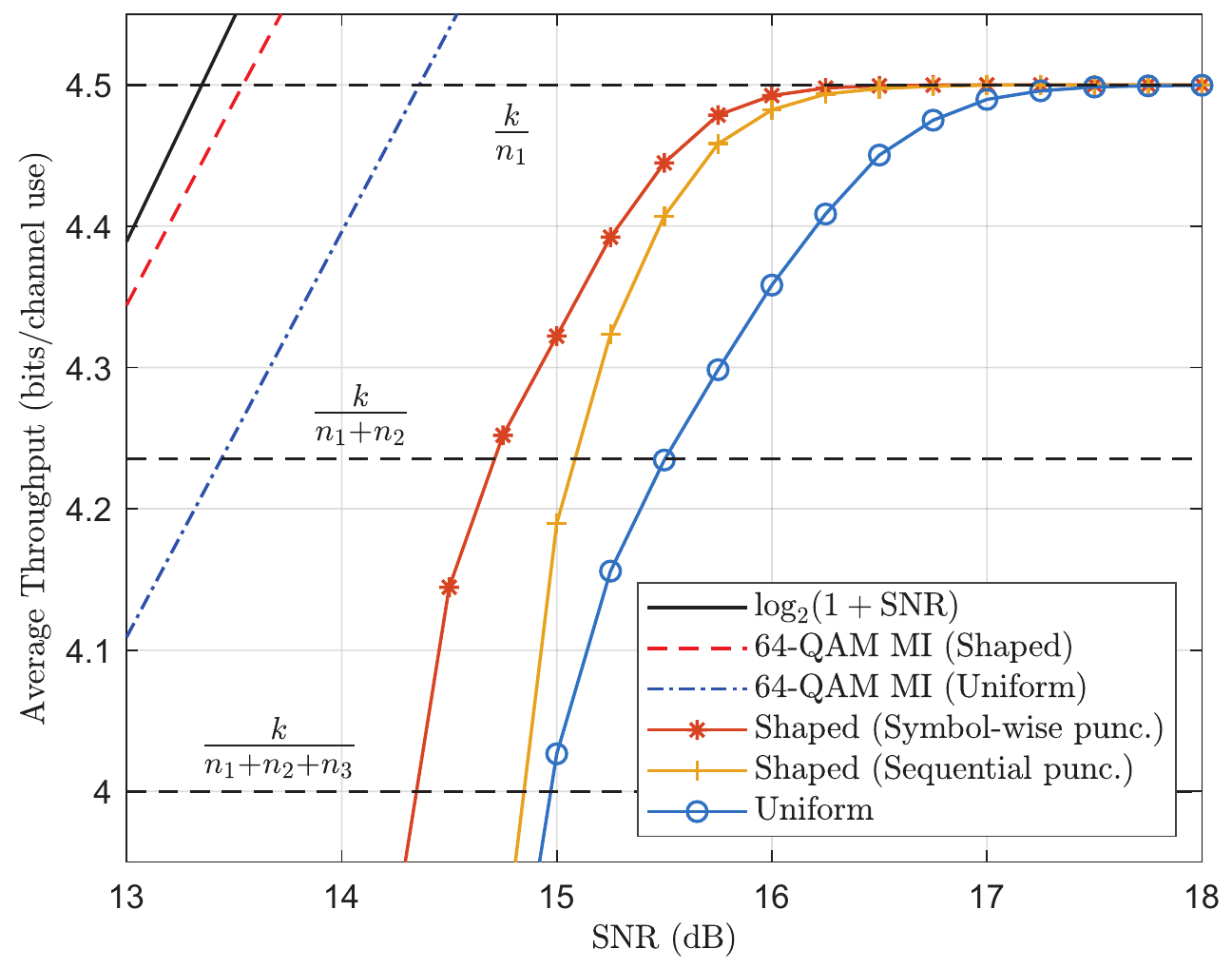}
        \caption{Average throughput performance for $k = 864$, $t_\text{max}=3$, and $(n_1,n_2,n_3)=(192,12,12)$.}
        \label{fig:5}
\end{figure}

\begin{figure}[!t]
        \centering
        \includegraphics[scale=0.6]{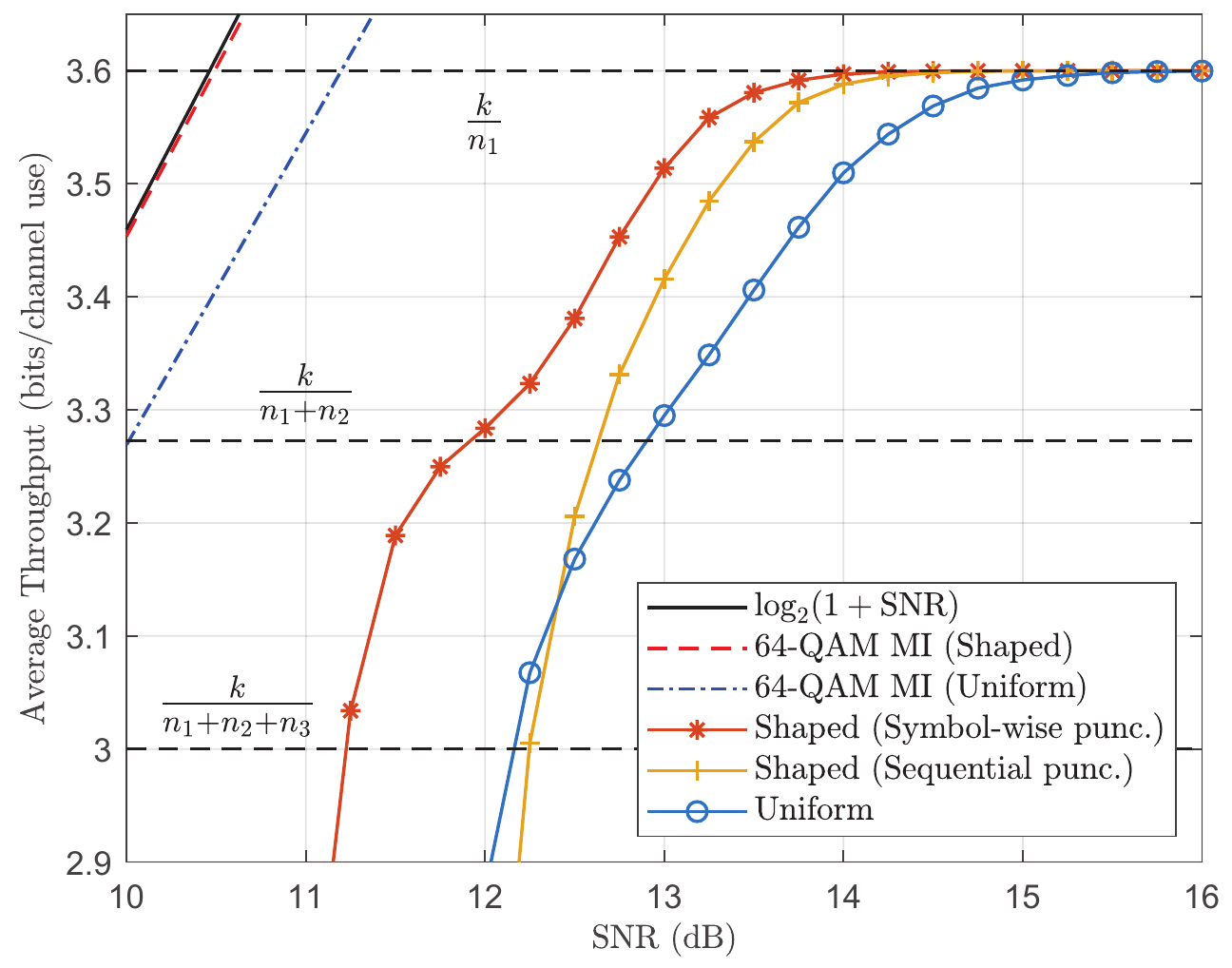}
        \caption{Average throughput performance for $k = 648$, $t_\text{max}=3$, and $(n_1,n_2,n_3)=(180,18,18)$.}
        \label{fig:6}
\end{figure}

We start with the throughput performance. The average throughput is calculated by
\begin{equation}
        \setlength{\abovedisplayskip}{4pt}
        \setlength{\belowdisplayskip}{4pt}
        \text{TP} = \sum_{t=1}^{t_\text{max}} \frac{k}{\sum_{j=1}^t n_j} \cdot \Pr\left\{ \mathcal{E}_t \right\},
\end{equation}
where $\mathcal{E}_t$ denotes that the $t$-th transmission is successful while the first $t-1$ transmissions are failed. We evaluate the throughput performance of the symbol-wise puncturing over the AWGN channel with the data block length $k = 648$ and $864$ and the LDPC encoded codeword length $n_c = 1296$. Each codeword can be modulated to $n=216$ 64-QAM symbols and is allowed to be transmitted a maximum of $t_\text{max}=3$ times. More simulation parameters are detailed in Table~\ref{tab:1}. Fig.~\ref{fig:5} and Fig.~\ref{fig:6} show the average throughput as a function of SNR with $k = 864$ and $648$, respectively. For comparison, we evaluate the average throughput of transmitting uniform and shaped bits with sequential puncturing. The Gaussian limit and the mutual information (MI) with uniform and shaped 64-QAM inputs are also provided. We observe that the shaped symbol-wise puncturing achieves a stable shaping gain across the SNR of at least 0.6 dB compared with the uniform under the same throughput. Note that the gain remains almost a constant with the increase of retransmission times. Instead, the gain of the shaped sequential puncturing drops rapidly when retransmission is required. In other words, at the same SNR, our symbol-wise puncturing can bring higher average throughput.

\begin{table}[tb]
        \caption{Average Throughput Simulation Parameters}
        \label{tab:1}
        \centering
        \renewcommand\arraystretch{1.05}
        \begin{tabular}{ccc}
        \hline
        Parameter                                                         & \multicolumn{2}{c}{Value} \\ \hline
        $k$                                                               & 648         & 864         \\
        Modulation                                                        & 64-QAM      & 64-QAM      \\
        $t_\text{max}$                                                    & 3           & 3           \\
        $n$                                                               & 216         & 216         \\
        $(n_1, n_2, n_3)$                                                 & (180, 18, 18) & (192, 12, 12) \\
        \begin{tabular}[c]{@{}c@{}}LDPC code rate\\ (Uniform)\end{tabular} & 648/1296    & 864/1296    \\
        \begin{tabular}[c]{@{}c@{}}LDPC code rate\\ (Shaped)\end{tabular}  & 972/1296    & 1080/1296   \\
        $k^\prime$                                                        & 590         & 700         \\
        $R_{dm}$                                                          & 1.3675      & 1.6218      \\
        \begin{tabular}[c]{@{}c@{}}Desired \\ distribution\end{tabular} &
        \begin{tabular}[c]{@{}c@{}}(0.5764, 0.3148, \\ \enspace 0.0926, 0.0162)\\ (Optimized for 8.5 dB)\end{tabular} &
        \begin{tabular}[c]{@{}c@{}}(0.4792, 0.3241, \\ \enspace 0.1505, 0.0463)\\ (Optimized for 12 dB)\end{tabular} \\ \hline
        \end{tabular}
\end{table}

\begin{table}[tb]
        \caption{Average Distribution of Each Transmission with $k=648$\\ (Statistics at 11 $\rm{dB}$).}
        \label{tab:2}
        \centering
        \renewcommand\arraystretch{1.05}
        \begin{tabular}{lc}
        \hline
        \multicolumn{2}{c}{Symbol-wise puncturing}    \\ \hline
        Desired $P_A$         & (0.5764, 0.3148, 0.0926, 0.0162) \\
        Average $P_A$(1st Tx) & (0.5763, 0.3149, 0.0926, 0.0162) \\
        Average $P_A$(2nd Tx) & (0.5769, 0.3139, 0.0929, 0.0163) \\
        Average $P_A$(3rd Tx) & (0.5787, 0.3130, 0.0928, 0.0156) \\ 
        Average $P_A$(all Txs)& (0.5765, 0.3147, 0.0926, 0.0162) \\ \hline\hline
        \multicolumn{2}{c}{Sequential puncturing}     \\ \hline
        Desired $P_A$         & (0.5764, 0.3148, 0.0926, 0.0162) \\
        Average $P_A$(1st Tx) & (0.5763, 0.3149, 0.0926, 0.0162) \\
        Average $P_A$(2nd Tx) & (0.2494, 0.2511, 0.2505, 0.2489) \\
        Average $P_A$(3rd Tx) & (0.2057, 0.2788, 0.3070, 0.2085) \\
        Average $P_A$(all Txs)& (0.5198, 0.3067, 0.1227, 0.0508) \\ \hline
        \end{tabular}
\end{table}

\begin{figure}[!t]
        \centering
        \includegraphics[scale=0.6]{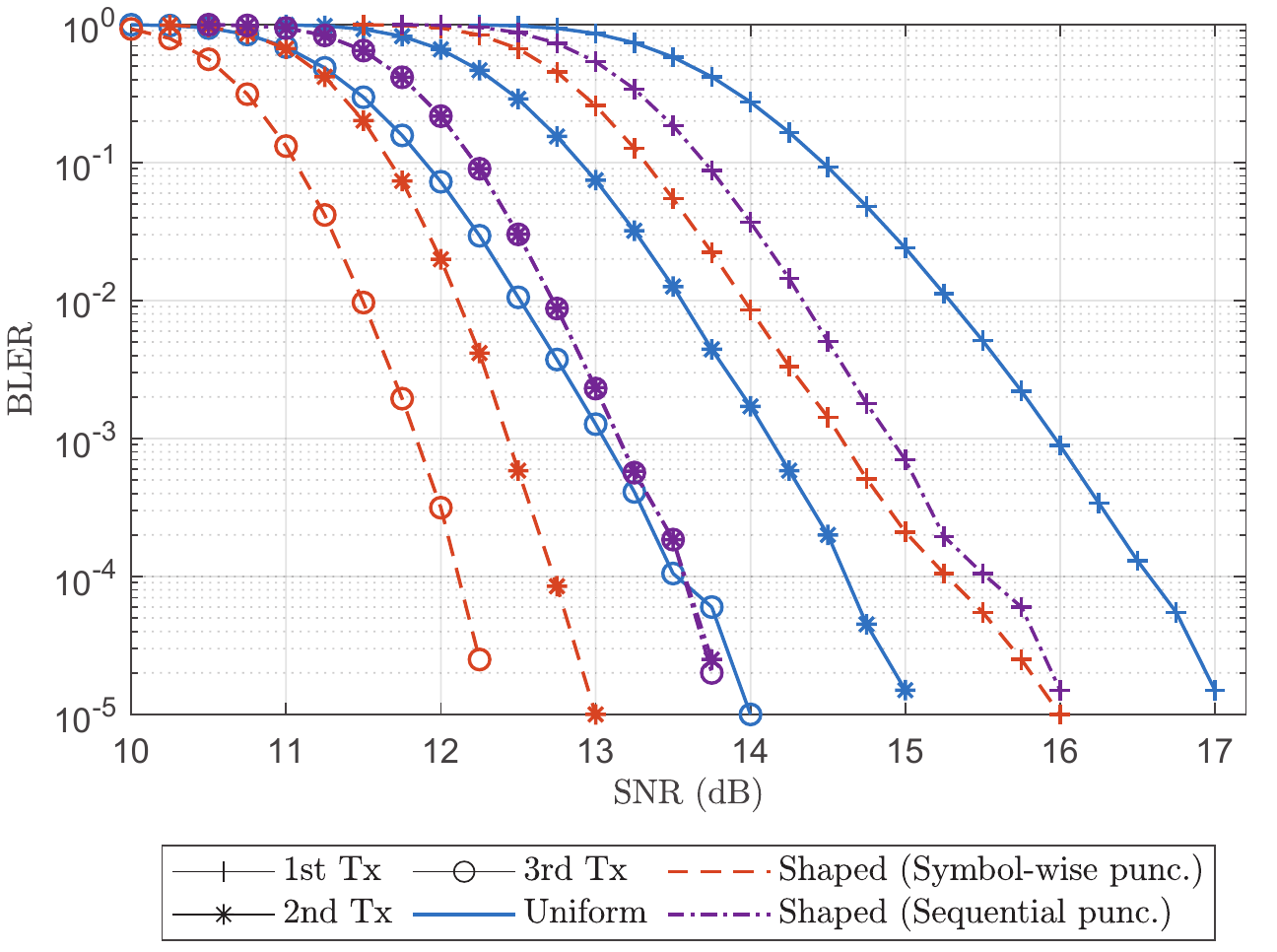}
        \caption{BLER performance of each transmission with $k=648$.}
        \label{fig:7}
\end{figure}

To futher illustrate the performance of symbol-wise puncturing, we average the probability distribution of each transmission with $k=648$ after transmitting 10,000 data blocks, as shown in Table~\ref{tab:2}. The block error rate (BLER) of each transmission is also plotted in Fig.~\ref{fig:7}. We observe that our symbol-wise puncturing preserves the desired distribution with only a negligible distortion during each transmission. However, in retransmissions, the average distribution of the shaped sequential puncturing tends to be uniform, which degrades some shaping gain. This can be verified in Fig.~\ref{fig:7}. At a BLER of $10^{-3}$, the symbol-wise puncturing achieves over 1.2 dB gain per transmission compared with the uniform case. The shaped sequential puncturing achieves diminishing gain in the first two transmissions and even has no enhancement on the BLER after the third transmission due to the distortion of distribution. Meanwhile, we observe that although the shaped sequential puncturing and symbol-wise puncturing have the same average distribution at the first transmission, the symbol-wise puncturing performs better on BLER. This is because the symbol-wise puncturing transmits more parity bits in the first transmission, while the punctured amplitude bits can be well recovered with the help of a-priori information.

\begin{figure}[!t]
        \centering
        \includegraphics[scale=0.6]{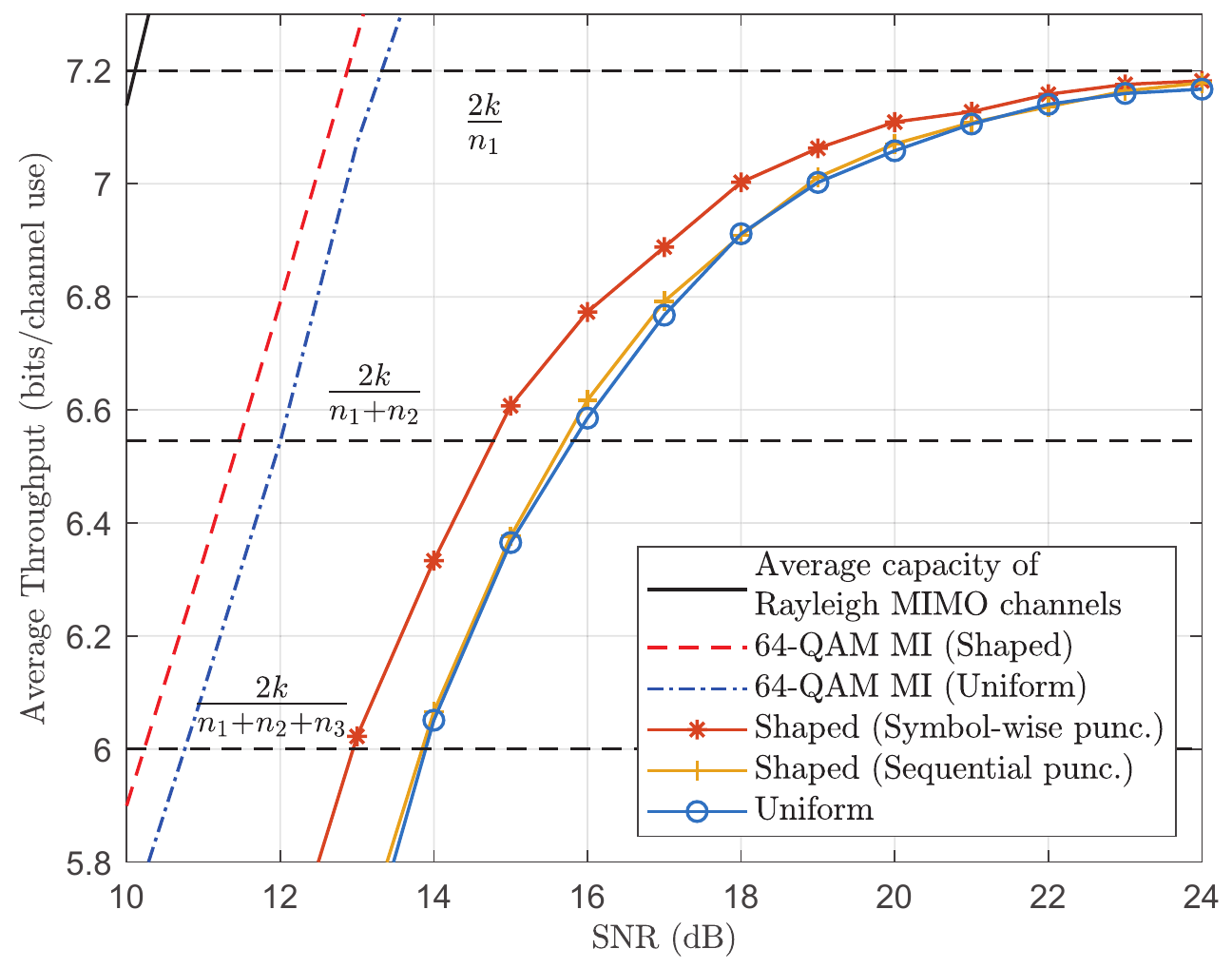}
        \caption{Average throughput performance for $k = 648$ over $2\times 2$ Rayleigh fading channels.}
        \label{fig:8}
\end{figure}

We also extend our work to MIMO fading channels. Fig.~\ref{fig:8} shows the average throughput performance with $k = 648$ over $2\times 2$ Rayleigh fading channels. We apply the non-iterative maximum likelihood detector at the receiver and other simulation parameters are the same as those in Fig.~\ref{fig:6}. The results indicate that the proposed symbol-wise puncturing still achieves a stable shaping gain across the SNR of about 1 dB, while the sequence puncturing achieves no gain at all.

\section{Conclusion}
In this letter, we proposed a symbol-wise puncturing scheme for IR-HARQ integration with PAS, which performs symbol-wise puncturing on the label sequence of the shaped modulation symbols to preserve the probability distribution per transmission. Simulation results showed that the symbol-wise puncturing scheme not only achieves stable shaping gain across the SNR, but also ensures the desired probability distribution per transmission. In the future, our work can focus on the possible extension to more HARQ schemes and FEC codes.


\vfill

\end{document}